\documentclass{article}
\usepackage{graphicx,floatrow}
\usepackage{dcolumn}
\usepackage{bm, authblk}

\usepackage{amsmath, amssymb, amsfonts}
\newtheorem{theorem}{Theorem}[section]

\newtheorem{rem}[theorem]{Remark}

\newcommand{\beq}{\begin{equation}}
\newcommand{\eeq}{  \end{equation}}
\newcommand{\beqa}{\begin{eqnarray}}
\newcommand{\eeqa}{  \end{eqnarray}}

\newcommand{\grad}{{\boldsymbol \nabla}}
\def\strutdepth{\dp\strutbox}
\def\nw#1{\strut\vadjust{\kern-\strutdepth\vtop to0pt{\vss\hbox to\hsize
{\hskip\hsize\hskip5pt$\leftarrow$\hss\strut}}}{\em #1}}

\begin{document}

\title{Spatial structure of shock formation}

\author[1]{J. Eggers}
\author[1,2]{T. Grava}
\author[3,1]{M.A. Herrada}
\author[2]{G. Pitton}
\affil[1]{School of Mathematics, 
University of Bristol, University Walk,
Bristol BS8 1TW, United Kingdom }
\affil[2]{SISSA, Via Bonomea 265, I-34136 Trieste, Italy}
\affil[3]{E.S.I., Universidad de Sevilla, Camino de los Descubrimientos 
s/n 41092, Spain }

\maketitle

\begin{abstract}
The formation of a singularity in a compressible gas, as described by 
the Euler equation, is characterized by the steepening, and eventual
overturning of a wave. Using a self-similar description in two space
dimensions, we show that the spatial structure of this 
process, which starts at a point, is equivalent to the formation of a 
caustic, i.e. to a cusp catastrophe. The lines along which the profile 
has infinite slope correspond to the caustic lines, from which we 
construct the position of the shock. By solving the similarity equation, 
we obtain a complete local description of wave steepening and of the 
spreading of the shock from a point. 
\end{abstract}

\section{Introduction}
From well into the 19th century, it has been known that the 
equations of compressible gas dynamics form shocks, i.e. 
lines or surfaces across which variables change in a discontinuous 
fashion (\cite{Courant_Friedrichs,LL84a}). This makes them perhaps 
the earliest example of a singularity of solutions to a partial 
differential equation (\cite{EF_book}). For smooth initial data, 
shock formation is associated with 
a gradual steepening, and eventual overturning of the velocity and 
density profiles. A shock develops  at the point where the slope first 
becomes infinite. The shock location can be calculated from the 
overturned profile via the so-called Rankine-Hugoniot conditions 
(\cite{Courant_Friedrichs}). The generic solution of hyperbolic 
(not linearly degenerate) systems in one space dimension with smooth 
initial data develops a cusp catastrophe, while solution to elliptic 
systems in one space dimension develop an elliptic umbilic catastrophe 
(\cite{DGKM15}).

\begin{figure}
\centering
   \includegraphics[width=0.9\textwidth]{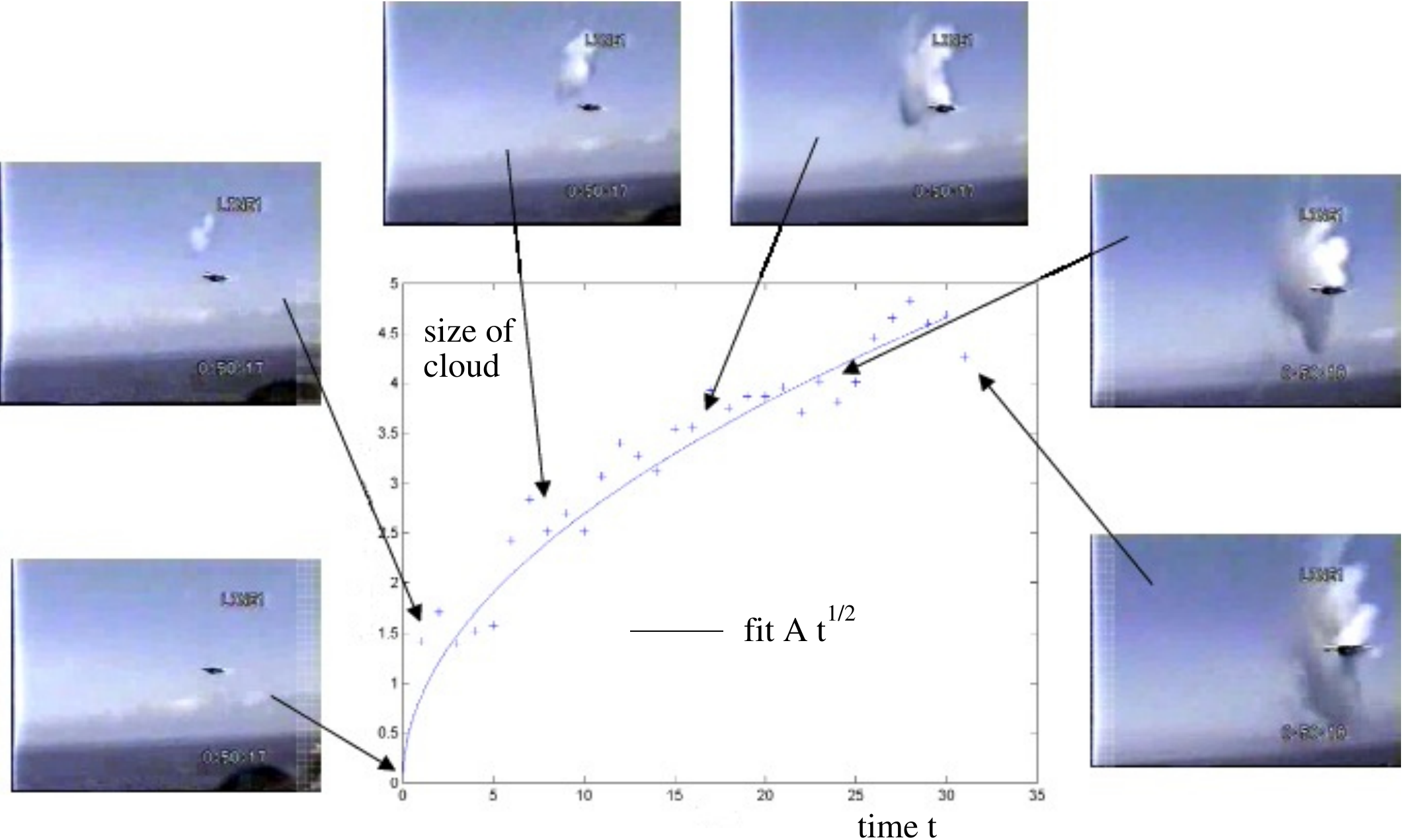} 
  \caption{
The spreading of a shock wave behind a supersonic plane, as 
marked by the condensation cloud produced by the shock. 
The data are based on measurements from a video 
(https://www.youtube.com/watch?v=gWGLAAYdbbc), with some sample 
images shown. Image and data analysis by Patrice Legal, used
with permission. The width of the cloud scales like $t^{1/2}$,
as measured from the initiation of the cloud. Absolute units of 
space and time are unknown.
}
 \label{fig:plane}
   \end{figure}

Relatively little emphasis has been placed on the description of 
how a shock is formed initially, starting from smooth initial data. 
The expectation is that the solution near the singular point is 
self-similar (\cite{EF_book}), but self-similar properties, in particular 
in more than one dimension, have also not received much attention until 
recently (\cite{PLGG08,EF09,MS08,MS12}).
In the transversal direction, the size of the shock solution scales 
like the square root of time, a fact which is confirmed readily from 
observation, see Fig.~\ref{fig:plane}. 

It has been conjectured for a long time (\cite{T76,PS78}) that the formation 
of a shock  in gas dynamics is analogous to the formation of caustics of 
wave fields 
(\cite{Nye99}), and thus are part of the same hierarchy of singularities
which can be classified using catastrophe theory 
(\cite{Berry81,Arnold89,Arnold90b}). The simplest such singularity is the 
fold, which originates from a point of higher symmetry called the 
cusp catastrophe (\cite{Nye99}). Thus the cusp catastrophe is the point 
where the singularity is expected to occur for the first time, unless
initial conditions are chosen such that the catastrophe is of higher order
(\cite{Nye99}). 
Examples of experimental observations of cusp catastrophes are found in optics 
(\cite{Nye99}), shock waves (\cite{SK76}), and clouds of cold atoms 
(\cite{RBSKARD14}). Note however that the cusp catastrophe considered 
for example in (\cite{SK76,CS78,CC90,CS97a}) appears in the shape of the 
shock front itself, whereas we consider the evolution of the velocity 
and density fields as a shock is formed. 

In order to use catastrophe theory, one needs to describe the phenomenon 
by means of a smooth mapping, whose singularities can be classified.
In optics, Fermat's principle guarantees the existence of 
such a function (\cite{Nye99}). In the case of shock dynamics, the method 
of characteristics can provide an analogous function (\cite{Arnold89}), but 
its existence is usually guaranteed only
in one space dimension (\cite{Courant_Friedrichs}) or for the simplest 
purely kinematic equation (\cite{PLGG08}). In \cite{GKE16} we proposed 
an extension of the method of characteristics for the dKP equation 
(\cite{GKE16}), which removes the singularity in the neighborhood of a shock, 
so that the unfolded profile can be expanded about the shock position. 
However, in the case of the full two or three-dimensional equations of 
compressible gas dynamics, no such smooth unfolding is known to exist, so 
catastrophe theory or an analogous method of expansion cannot be applied. 

Instead, we resort to solving the equations of motion directly near
the singularity, whose structure is expected to resemble the cusp 
catastrophe of geometrical optics. The key idea is to use the self-similar
properties of the cusp catastrophe, in order to obtain a leading-order 
solution of the equations of motion in powers of the time distance 
$t' = t_0 - t$ to the singularity, where $t_0$ is the time of blow-up. 

\section{Equations of motion}

We consider the compressible Euler equation in two space dimensions, 
and denote the spatial variables by $\boldsymbol{x}=(x,y)\in\mathbb{R}^2$. 
The velocity field ${\bf v}=(u,v)$ is assumed irrotational: 
${\bf v} =  \nabla\phi$. Before the formation of a shock, we can consider 
the flow to be isentropic. For simplicity, we assume the relation 
between density $\rho$ and pressure $p$ to be described by the polytropic 
ideal gas law (\cite{LL84a})
\beq
p = \frac{A}{\gamma} \rho^{\gamma}. 
\label{adiabatic}
\eeq
The compressible Euler system consists of three equations for the functions
$\rho$ and ${\bf v}=(u,v)$, which correspond to balance statements for 
mass and linear momentum:
\beq
\frac{\partial \rho}{\partial t} + \nabla\cdot(\rho{\bf v}) = 0,
\label{continuity}
\eeq
\beq
\label{Euler}
\frac{\partial {\bf v}}{\partial t} + ({\bf v}\cdot\nabla){\bf v} 
= -\frac{1}{\rho}{\bf \nabla}p ;
\eeq
here $\nabla=(\partial_x,\partial_y)$.
Using the potential flow assumption, (\ref{Euler}) can be integrated to 
\beq
\label{Ber}
\frac{\partial \phi }{\partial t}+\frac{1}{2}\left|\nabla\phi\right|^2
= -\frac{A}{\gamma-1}\left(\rho^{\gamma-1} - \rho_0^{\gamma-1}\right).
\eeq

Note that we have the freedom to add an arbitrary constant 
on the right-hand side, which can be absorbed into the potential 
with the transformation 
$\phi \to \phi + \frac{A}{\gamma-1}\rho_0^{\gamma-1}t$. Here we 
choose it such that the right-hand side vanishes at the position 
of the shock. 

The isentropic compressible Euler equation admits classical 
solution if the initial data is sufficiently regular (\cite{Lax72}).
However it is well-known that, even starting from extremely regular 
initial data, the solution develops singularities in  finite  time 
(\cite{Majda84}, \cite{Christo}). An estimate of the blow-up time  of 
classical solutions has been obtained in (\cite{Alinhac93}) for small 
perturbations of constant initial data.

In this manuscript we address the nature of singularity formation 
for classical solutions. After the formation of the singularity the 
solution exists only in a weak sense, and hence to fix a solution uniquely,
extra conditions have to be imposed. When dealing with systems 
coming from physics,  the second law of thermodynamics naturally induces 
such conditions, by assuming that weak solutions satisfy certain entropy 
inequalities (which correspond to the Rankine-Hugoniot  conditions
(\cite{LL84a})).
The theory is quite mature  for hyperbolic systems  in one 
space dimension or for hyperbolic scalar equations in more then one 
space dimension. In these cases  the Rankine-Hugoniot conditions single 
out uniquely a solution which coincides with that obtained in the limit
of vanishing viscosity, see e.g. (\cite{Kruzkov}), 
(\cite{Bressan2005}).

When dealing with systems of conservation laws in more than one space 
dimension, it is still an intriguing mathematical problem to develop a 
theory of well-posedness for the Cauchy problem which includes the 
formation and evolution of shock waves. In particular for the compressible 
Euler equation in two space dimensions it has been shown that the entropy 
inequalities do not guarantee uniqueness and some counter-examples 
are obtained for initial data that are locally Lipschitz
(\cite{CDL15}, \cite{Elling06}). However in this manuscript we are 
interested in the evolution of a classical solution (at least $C^1$) 
into  its  first singularity, and to the local structure of the shock 
near this singularity.
 
 Below we will consider the coupled set of equations 
(\ref{continuity}),(\ref{Euler}). Since entropy is created in a shock,
the adiabatic gas law (\ref{adiabatic}) and thus (\ref{Euler}) will 
no longer strictly be valid after shock formation. However, for a short 
time entropy production is still weak, so we will 
still be able to use an adiabatic description to leading order. 

\section{Similarity structure}
\begin{figure}
\centering
   \includegraphics[width=0.9\textwidth]{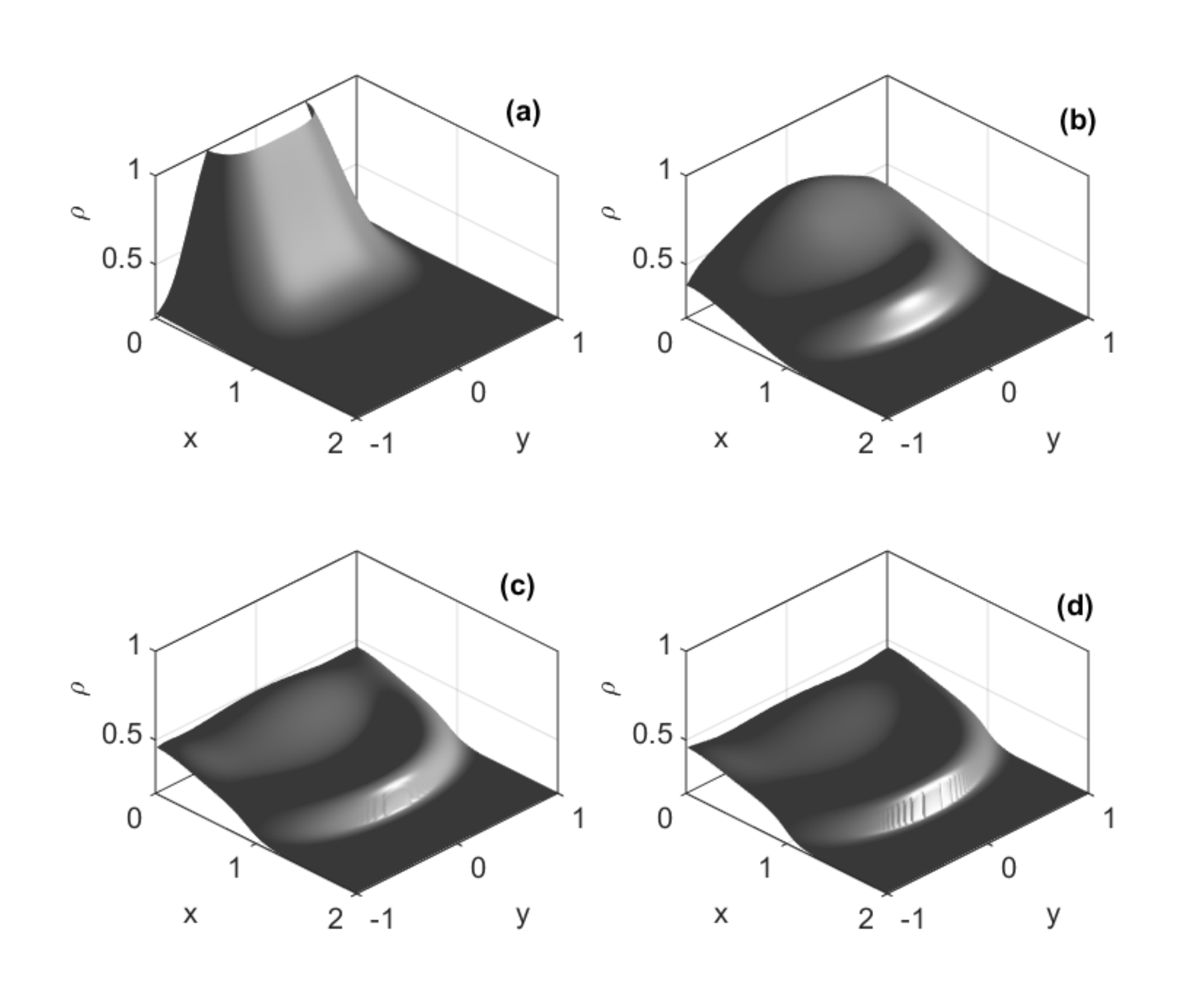} 
  \caption{
Time evolution of the density, as described by the compressible 
Euler equation at $t=0$ (a), $t=0.4$ (b), $t=0.511$ (c), and $t=0.55$ (d).
The initial condition is a concentrated density in an initially quiescent 
fluid, as given in (\ref{init}). At (c), a shock forms, which has 
spread in (d). 
}
 \label{fig:overview}
   \end{figure}
We are interested in describing the formation of a singularity in 
solutions of the compressible Euler equation. At the point  where the 
singularity first forms, the gradients of all variables $\rho, u, v$ 
blow up, while the variables themselves remain finite. 
In the generic case, the singularity develops at a point 
(the gradient blowing up along a line corresponds to a non-generic initial 
condition); we denote  the conditions at this point (such as the density 
$\rho_0$ or the velocity $\boldsymbol{v}_0=(u_0,v_0)$) with the subscript zero. 
We assume that at the critical time $t_0$ the gradient $\nabla u$ blows up 
at one point $(x_0,y_0)$ in all directions of the $(x,y)$ plane except one,
in which it remains bounded. By contrast, a gradient blowing up in all 
directions corresponds to an elliptic umbilic singularity, typical of 
elliptic systems. 

Using the invariance of the Euler equation under rotation 
in the $(x,y)$- plane, we denote the direction where the gradient of $u$ 
remains  bounded at the critical point by $y$, while 
$\partial x/\partial u=0$. Since the flow is potential, it follows that 
the first derivative  $v_x=u_y$  remains bounded at the singular point.
The condition that the profile has not already overturned amounts to 
demanding that $\partial^2 x/\partial u^2 = 0$, while the
third derivative will in general be finite (\cite{LL84a}, \cite{Majda84}). 
Thus in summary at the point of the wave profile first becoming singular 
we have the conditions
\beq
\frac{\partial u}{\partial y} = const, \quad
\frac{\partial x}{\partial u} = 0, \quad
\frac{\partial^2 x}{\partial u^2} = 0, \quad
\frac{\partial^3 x}{\partial u^3} = const.  
\label{shock_cond}
\eeq
This is illustrated in Fig.~\ref{fig:overview}, which shows 
an example of a numerical simulation of the Euler equation to be 
described in more detail in Section~\ref{sec:num}. It starts from a 
smooth initial condition for the density, whose profile gradually 
steepens, until a shock is formed at a point $(x_0,y_0=0)$ at the 
time $t = t_0$ (panel (c)). For $t>t_0$, the shock spreads along a 
line transversal to the direction of propagation (in the $x$-direction),
while the height of the jump increases. 

We move into a frame of reference such that \[
{\bf v}_0\equiv(u_0,v_0) = 0
\]
at the point where the singularity  is formed. 
The speed of sound at the singular point is 
\beq
c_0^2 = \frac{\partial p}{\partial \rho} = A \rho_0^{\gamma-1}.
\label{sound}
\eeq
 To describe the neighborhood of the singularity, 
we use a self-similar description (\cite{EF_book}), in analogy to caustic 
singularities in two dimensions (\cite{EHHS14}), and shocks in the 
dKP equation (\cite{GKE16}). 

In the self-similar region, we assume the scalings
$x' \propto t'^{\beta_1}$, $y'\propto t'^{\beta_2}$, and 
$u \propto t'^{\alpha}$, where $x' = x - x_0$, $y' = y - y_0$, 
and $t' = t_0 - t$, so that $t'>0$ before the singularity, and 
$t'<0$ after. Balancing the first two terms in the Euler 
equation (\ref{Euler}) in the propagation direction, we obtain 
$\alpha - 1 = 2\alpha - \beta_1$. If in analogy to (\ref{shock_cond})
we demand $\partial^3 x / \partial u^3 = const$, we have
$3\alpha = \beta_1$, so that $\alpha = 1/2$ and $\beta_1 = 3/2$. 
Moving in the transversal ($y$) direction, the blow-up of the gradient 
will occur at a slightly later time 
$t_c(y') - t_0 = a y'^2 + O(y^3)\propto t'$; the linear term must 
vanish, since otherwise there would be a $y'\ne 0$ where blow-up 
would occur at a time earlier than $t_0$. Hence it follows that 
$\beta_2 = 1/2$. The scaling exponents correspond
to those found previously for wave breaking (\cite{PJLLA08},\cite{PLGG08}),
the cusp caustic (\cite{EHHS14}), and
for shock formation in two dimensions (\cite{GKE16}). 

Since the shock travels on the back of a sound wave with speed $c_0$
in the $x$-direction, we consider the ansatz
\beq
\begin{split}
&\phi(x,y,t) =|t'|g_2(\eta)+|t'|^{\frac{3}{2}}g_1(\eta) + 
|t'|^2\Phi(\xi,\eta)+|t'|^{\frac{5}{2}}\Phi_1(\xi,\eta)+\dots \\
&\xi = \frac{x' + c_0t' -c_1y' -By'^2}{|t'|^{3/2}},
\quad \eta = \frac{y'}{|t'|^{1/2}} 
\label{similarity_phi}
\end{split}
\eeq
for the potential. Observe  that 
\begin{equation}
\label{u_exp}
u(x,y,t) = \phi_x(x,y,t) = |t'|^{\frac{1}{2}}\Phi_{\xi}(\xi,\eta)+O(|t'|)
:=|t'|^{\frac{1}{2}}U(\xi,\eta)+O(|t'|)
\end{equation}
\[ v=\phi_y=|t'|^{\frac{1}{2}}(g_{2\eta}-c_1\Phi_{\xi}) + 
|t'|(g_{1\eta}-2B\eta\Phi_{\xi})+O(|t'|^{\frac{3}{2}}).\]
As in (\cite{EHHS14}), $-c_1y'-By'^2$ in $\xi$ is  a lower order term, 
which describes a modulation in the transversal direction. A third order 
term in $y$  would be proportional to $\eta^3$, which is already accounted 
for in the $\eta$ dependence of $\Phi$. The absolute sign guarantees that 
(\ref{similarity_phi}) works both before and after the singularity. 
For the density we make the ansatz 
\beq
\rho(x,y,t)=  \rho_0\left[ 1 + |t'|^{1/2}R(\xi,\eta) + 
|t'|Q(\xi,\eta) \right]+ O(t'^{3/2}),
\label{similarity_density}
\eeq
which solves (\ref{continuity}) and (\ref{Ber}) to leading order, 
as we will see now. The higher order contributions $\Phi_1$ and $Q$
are needed for consistency, but we will not calculate them here. 

Inserting (\ref{similarity_phi}),(\ref{similarity_density}) into 
(\ref{Ber}), we obtain 
\begin{equation}
\begin{split}
&\pm(\dfrac{1}{2}g_{2\eta}\eta-g_2)+
|t'|^{1/2}\left(\pm\dfrac{1}{2}(g_{1\eta}\eta- 3g_1)-c_0\Phi_{\xi}\right) +\\
& |t'|\left[ \mp 2\Phi \pm \frac{3\xi}{2} \Phi_{\xi} 
\pm \frac{\eta}{2} \Phi_{\eta} + \frac{1}{2}\Phi_{\xi}^2 + 
\dfrac{1}{2}(g_{2\eta}-c_1\Phi_{\xi})^2-c_0\Phi_{1\xi}\right]  = \\
&-c_0^2\left\{ |t'|^{1/2} R + |t'|[Q + \frac{1}{2}(\gamma-2)R^2] \right\} + 
O(t'^{3/2}).
\label{Eu2}
\end{split}
\end{equation}
Thus at order $|t'|^0 $ and $|t'|^{\frac{1}{2}}$  we have 
\begin{align}
\label{g2}
&\dfrac{\eta}{2}g_{2\eta}=g_2,\\
&c_0 R = \Phi_{\xi}\mp\dfrac{1}{2c_0}(g_{1\eta}\eta- 3g_1) = 
U(\xi,\eta)\mp\dfrac{1}{2c_0}(g_{1\eta}\eta- 3g_1).
\label{lead}
\end{align}
Equation (\ref{g2}) gives
\[
g_2(\eta)=a_0\eta^2,
\]
for some constant $a_0$. Since we expect the leading order term, 
$R(\xi,\eta)$, of $\rho$ to be continuous in the transversal direction 
$y$ near the singularity point,  we infer from (\ref{lead})  
that 
\beq
\label{cond_g}
 g_{1\eta}\eta- 3g_1 =0,
\eeq
so that one has 
\beq
\label{UR}
c_0 R = \Phi_{\xi} = U(\xi,\eta),\quad g_1(\eta)=a_1\eta^3
\eeq
for some constant $a_1$. Finally grouping together terms of order 
$|t|'$ in (\ref{Eu2})  and using (\ref{UR}), we obtain 
\beq
c_0^2 Q = \pm 2\Phi \mp \frac{3\xi}{2} \Phi_{\xi} \mp \frac{\eta}{2} \Phi_{\eta}
+ \frac{1}{2}(1-\gamma)\Phi_{\xi}^2-\dfrac{1}{2}(g_{2\eta} - 
c_1\Phi_{\xi})^2+c_0\Phi_{1\xi}\,.
\label{next}
\eeq

Next, inserting (\ref{similarity_phi}),(\ref{similarity_density}) into 
(\ref{continuity}), we have
\begin{align*}
&-c_0R_{\xi} |t'|^{-1} + |t'|^{-1/2}\left[\mp \frac{R}{2}
\pm\frac{3\xi}{2} R_{\xi} \pm\frac{\eta}{2} R_{\eta} - c_0 Q_{\xi}\right]
+ |t'|^{-1} (1+c_1^2)\Phi_{\xi\xi} \\
&+ |t'|^{-1/2}\left[\Phi_{\xi}R_{\xi} 
+ \Phi_{\xi\xi}R +4c_1B\eta\Phi_{\xi\xi} - 
c_1(g_{2\eta}-c_1\Phi_{\xi})R_{\xi}+\Phi_{1\xi\xi}\right] = O(t'^0),
\end{align*}
whose leading order part is compatible  with (\ref{lead}) if
\beq
c_1=0.
\eeq
 The next order, combined with (\ref{lead}), gives 
\beq
\label{Q2}
c_0^2Q_\xi=\mp \frac{\Phi_\xi}{2}
\pm\frac{3\xi}{2} \Phi_{\xi\xi} \pm\frac{\eta}{2} \Phi_{\xi\eta} +
2\Phi_{\xi}\Phi_{\xi\xi}+c_0\Phi_{1\xi\xi}\,.
\eeq

Differentiating  (\ref{next}) with respect to $\xi$ and subtracting 
(\ref{Q2}) one obtains
\beq
\Phi_{\xi} - 3\xi \Phi_{\xi\xi} - \eta \Phi_{\eta\xi} = 
\pm (\gamma+1)\Phi_{\xi}\Phi_{\xi\xi},
\label{sim_orig}
\eeq
which is a closed equation for $\Phi$. 
Summing (\ref{next}) with the integral of (\ref{Q2}) with respect to 
$\xi$ results in
\beq
c_0^2 Q = \frac{3-\gamma}{4} U^2-\dfrac{a_0}{2}\eta^2+2c_0\Phi_{1\xi},
\label{Q}
\eeq
where the scaling function $\Phi_1$ is left undetermined at the present 
level of approximation. Therefore, we have not made explicit 
the constant of integration in (\ref{Q}). 

Using (\ref{UR}) and the above relation we can express the density $\rho$ in 
(\ref{similarity_density})  has 
\beq
\rho = \rho_0\left[ 1 + \frac{|t'|^{1/2}}{c_0}U(\xi,\eta) + 
\frac{3-\gamma}{4 c_0^2}|t'|U^2(\xi,\eta) \right] +
|t'|\rho_0\left[2c_0\Phi_1(\xi,\eta)-\dfrac{a_0}{2}\eta^2 \right]+O(t'^{3/2}).
\label{rho}
\eeq
Note that this means that 
\beq
\rho = \rho_0\left[ 1 + \frac{\gamma-1}{2c_0} u\right]^{2/(\gamma-1)} +O(|t'|),
\label{simple}
\eeq
which, up to terms of order  $O(|t'|),$ is the form of a simple wave 
(\cite{LL84a}) for the one-dimensional Euler system.
In the limit $\gamma\to 1$ one obtains 
\[
\rho=\rho_0e^{\frac{u}{c_0}}+O(|t|'),
\]
which is also  consistent with the form of a simple wave in the case $\gamma=1$.

The case of a Karman-Tsien gas \cite{BH93} $\gamma =-1$ is special. 
This is because the structure of the similarity 
solution is different, and (\ref{similarity_phi}) is to be replaced by 
$\xi = (x'-By'^2)/t'^{3/2}$. In this solution, the pressure is subdominant, 
so one is effectively solving the kinematic equation. 
\begin{rem}
After setting 
\[
h(\xi,\eta) = -\frac{\gamma+1}{2}\Phi(\xi,\eta)
\]
in equation (\ref{sim_orig}) and integrating in $\xi$  one obtains 
\begin{equation}
4h - 3\xi h_{\xi} - \eta h_{\eta} \pm h_{\xi}^2  = 0,
\label{eik_sim_equ}
\end{equation}
which is the similarity equation derived for solutions of the eikonal
equation (\cite{EHHS14}). 
\end{rem}

Finally to solve (\ref{sim_orig}), we put $U=\Phi_{\xi}$ to obtain 
\beq
\label{eqU}
U - 3\xi U_ {\xi} - \eta U_{\eta} = 
\pm (\gamma+1)U U_{\xi},
\eeq
which can be linearized by transforming to $\xi(U,\eta)$:
\beq
\xi_U U - 3\xi +\eta\xi_{\eta} = \pm(\gamma+1)U,
\eeq
with general solution 
\beq
\xi = \mp \frac{\gamma+1}{2}U - U^3F\left(\frac{\eta}{U}\right).
\label{gen_sol}
\eeq

The form of the function $F=F(z)$ is set by the requirement that 
the similarity profile must be regular. From (\ref{gen_sol}) we
find that 
\[
\frac{\partial^4 \xi}{\partial \eta^4} =
-\dfrac{1}{U}F^{(iv)}\left(\frac{\eta}{U}\right),
\]
so by putting $\eta = aU$ for $a\neq 0$,  and letting $U\rightarrow 0$, 
one needs to impose $F^{(iv)}(a)=0$  for any nonzero constant $a$.
Hence $F$ is a cubic  polynomial, namely 
\beq
F(z) = A_0 + A_1 z + A_2 z^2 + A_3 z^3,
\label{F_form}
\eeq
and the similarity profile is 
\beq
\xi = \mp\frac{\gamma+1}{2}U - A_0 U^3 - A_1 U^2\eta - A_2 U \eta^2 
- A_3 \eta^3,
\label{gen_sol1}
\eeq
for some constants $A_0,A_1, A_2$ and $A_3$.

In principle one could use different values of these  constants for 
$t'>0$ and $t'<0$. However, we observe  that for fixed values of $x'$ 
and $y'$ away from $(0,0)$  the local structure of the solution has 
to be single-valued as a function of $t'$ as $|t|'\to 0$. It 
follows that  $U(\xi,\eta)$ has to be a single-valued function of 
$\xi$ and $\eta$ as $\xi\to\infty$ and $\eta\to\infty$, which
is possible only if the constants $A_0,A_1,A_2$ and $A_3$ have the 
same values before and after the singularity. 

This completes the solution; constraints on
the coefficients $A_i$ are given in (\ref{pos}) below. It is easy 
to confirm that the conditions (\ref{shock_cond}) are satisfied. 
Note that (\ref{gen_sol1}) corresponds exactly to the generic form of a 
cusp singularity (\cite{EHHS14,GKE16}) also found in the catastrophe 
theory of optical caustics  (\cite{Nye99}). In particular, there are
no quadratic terms in the expansion. From the condition that there can
be no overturning of the profile before shock formation (upper sign),
we also deduce the condition $A_0 > 0$. 

To determine the coefficients $A_i$ in (\ref{gen_sol1}) numerically, 
as we will do below, it is useful to take third derivatives of $x$ with 
respect to $u$ and $y$. First, at constant $y$, we have 
\[
x_u = u_x^{-1}, 
\]
and thus 
\beq
x_{uu} = -\frac{u_{xx}}{u_x^3}, \quad
x_{uuu} = -\frac{u_{xxx}}{u_x^4} + 3\frac{u_{xx}^2}{u_x^5}.
\label{xuuu}
\eeq
According to the implicit function theorem, 
\[
x_y = -\frac{u_y}{u_x},
\]
while 
\[
\left.\frac{1}{\partial y}\right|_u f(x,y) = 
\left.\frac{1}{\partial y}\right|_u f(x(u,y),y) = -f_x\frac{u_y}{u_x}
+ f_y, 
\]
and thus 
\[
x_{yy} = \left(\frac{u_y}{u_x}\right)_x\frac{u_y}{u_x} - 
\left(\frac{u_y}{u_x}\right)_y = 2\frac{u_{xy}u_y}{u_x^2} - 
\frac{u_{yy}}{u_x} - \frac{u_{xx}u_y^2}{u_x^3} 
\]
and 
\beqa
\label{xuyy}
&& x_{uyy} = \left[2\frac{u_{xy}u_y}{u_x^2} - 
\frac{u_{yy}}{u_x} - \frac{u_{xx} u_y^2}{u_x^3}\right]_x u_x^{-1},  \\
\label{xuuy}
&& x_{uuy} = \left(\frac{u_{xx}}{u_x^3}\right)_x \frac{u_y}{u_x}
-\left(\frac{u_{xx}}{u_x^3}\right)_y, \\
\label{xyyy}
&& x_{yyy} = -\left[2\frac{u_{xy}u_y}{u_x^2} - 
\frac{u_{yy}}{u_x} - \frac{u_{xx} u_y^2}{u_x^3}\right]_x \frac{u_y}{u_x}
+\left[2\frac{u_{xy}u_y}{u_x^2} - 
\frac{u_{yy}}{u_x} - \frac{u_{xx} u_y^2}{u_x^3}\right]_y. 
\eeqa

Using the scaling (\ref{similarity_phi}), the derivatives can 
be converted to similarity variables, so from (\ref{gen_sol1}) one
obtains
\beq
A_0 \simeq  -\frac{x_{uuu}}{6}, \quad A_1 \simeq -\frac{x_{uuy}}{2}, \quad
A_2 \simeq -\frac{x_{uyy}}{2}, \quad A_3 \simeq -\frac{x_{yyy}}{6},
\label{coefficients}
\eeq
to be evaluated at the critical point $t=t_0$, $x=x_0$ and $y=y_0$. 
Here and below, we are assuming that the higher order scaling functions
which appear in (\ref{similarity_phi}) are regular, so that the 
higher order contributions to to $u$ in (\ref{u_exp}) become 
negligible near the critical point.

Finally, the constant $B$ in (\ref{similarity_phi}) can be evaluated 
by computing the second derivative with respect to $y$:
\[
u_{yy} \simeq 4B^2\eta^2|t'|^{-3/2} U_{\xi\xi} - 2B|t'|^{-1} U_{\xi}
+ |t'|^{-1/2} U_{\eta\eta}. 
\]
Thus at $\eta=\xi=0$,  using 
$U_{\xi} = \xi_{U}^{-1} = \mp2/(\gamma+1)$, $U_{\xi\xi}=0$  and $U_{\eta\eta}=0$,
one finds that 
\beq
B \simeq\frac{\gamma+1}{4} u_{yy}(t_0 - t)
\label{B}
\eeq
as $t'\rightarrow 0$. 
Summarizing, the solution  near the  singularity at the point 
$(x_0,y_0, t_0)$ in the physical variables $x'=x-x_0$, $y'=y-y_0$ and 
$t'=t_0-t$ takes the form
\beq
\label{ss_u}
x'-c_0t'-By'^2\simeq-\dfrac{\gamma+1}{2}t'u-A_0u^3-A_1u^2y'-A_2uy'^2-A_3y'^3,
\eeq
where the constant $A_0,A_1,A_2$ and $A_3$ are determined in 
(\ref{coefficients}) and the constant $B$ is determines in (\ref{B}).

\begin{rem}
We   claim that the local structure of the singularity for the velocity 
$u(x,y,t)$ is captured by the self-similar profile obtained in 
(\ref{ss_u}). This represents the  leading order  term in the multiple scale 
expansion of $u(x,y,t)=|t'|^{\frac{1}{2}}U(\xi,\eta)+O(|t'|)$.
We will support this claim by a numerical example  presented in 
Sect.~\ref{sec:num}.
Furthermore, if we assume that  the higher order corrections to the 
potential field are regular at the singular point, one can deduce that 
the gradient of $u$ with respect to $y$ is not only constant but 
zero at the singular point. This fact is certainly true for smooth initial 
data that are invariant with respect to the symmetry $y\to -y$. 
\end{rem}

\section{After the singularity}
After a shock occurs, the adiabatic law (\ref{adiabatic}) is 
no longer valid, since entropy is generated inside the 
shock front, the entropy being given by 
\beq
s = c_v\ln\frac{p}{\rho^{\gamma}},
\label{entropy}
\eeq
where $c_v$ is the specific heat, which for simplicity
we consider constant. However, the jump in entropy across 
the shock is only of order $|t'|^{3/2}$, which results in a 
subleading contribution to (\ref{Eu2}). Following (\cite{LL84a}),
and using 
\[
w = \frac{\gamma}{\gamma-1}\frac{p}{\rho}
\]
for the enthalpy
of a polyatomic gas, the Rankine-Hugoniot jump condition across
a shock in a frame of reference moving with the shock is 
\beq
\frac{\gamma}{\gamma+1}\left(\frac{p_1}{\rho_1}-\frac{p_2}{\rho_2}\right)
+ \frac{1}{2}\left(\frac{1}{\rho_1}+\frac{1}{\rho_2}\right)
\left(p_2 - p_1\right) = 0,
\label{RH}
\eeq
where the index 1 denotes the front of the shock, index 2 the back. 
Combining (\ref{entropy}) and
(\ref{RH}), and expanding in the size of the pressure jump 
$p_2 = p_1 + \Delta p$, one finds 
\beq
\Delta s \equiv 
s_2 - s_1 = \frac{c_v}{12}\frac{\gamma^2-1}{\gamma^2}\frac{\Delta p^3}{p_1^3}.
\label{s_jump}
\eeq
Thus the jump in entropy is only of third order, which a fact which 
remains true for a gas of arbitrary thermodynamic properties (\cite{LL84a}).

From the solution (\ref{similarity_density}), we have that 
$\Delta p \propto t'^{1/2}$, and so it follows that 
$\Delta s \propto t'^{3/2}$, which means that 
\[
p = \frac{A}{\gamma} \rho^{\gamma} + O(t'^{3/2}). 
\]
Clearly, this makes a contribution of order $t'^{3/2}$ to (\ref{Eu2}),
which can be neglected. 
Given the leading-order solution, one can use the entropy production 
(\ref{s_jump}) to calculate the distribution of entropy near the shock,
using the convection equation 
\beq
\left(\frac{p}{\rho^{\gamma}}\right)_t + {\bf v}\cdot\nabla
\left(\frac{p}{\rho^{\gamma}}\right) = 0,
\label{s_conv}
\eeq
which says that entropy is transported with each fluid element, 
but not produced outside of the shock. 

\subsection{Shock condition}
\begin{figure}
\centering
   \includegraphics[width=0.49\textwidth]{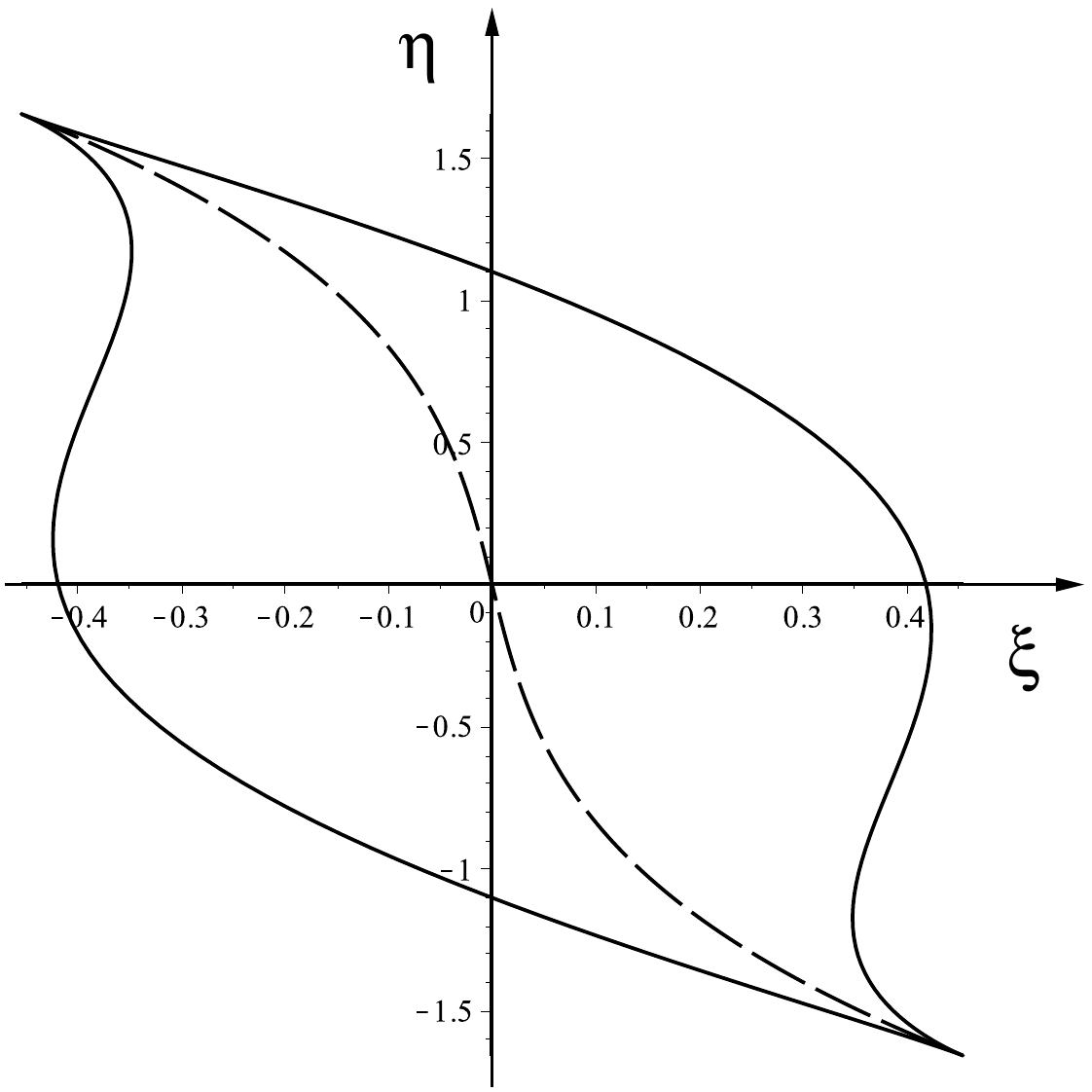} 
  \caption{The lip-shaped region inside which the velocity profile 
overturns; parameters are chosen arbitrarily as 
$(A_0,A_1,A2,A3) = (2,0.3,0.5,0.1)$ and $\gamma = 5/3$. The dashed line 
marks the position of the shock. 
}
 \label{fig:lip}
   \end{figure}
After the singularity, the solution given by (\ref{gen_sol1}) 
has a region where the profile has overturned. The line along 
which the profile is vertical is given by $\partial \xi/\partial U = 0$,
and thus for $t > t_0$:
\beq
\frac{\gamma+1}{2} = 3A_0 U^2 + 2A_1 U\eta + A_2 \eta^2 .
\label{lip}
\eeq
This can be parameterized as an ellipse in $(U,\eta)$-space, provided 
that the quadratic form on the right is positive definite; for this 
we need that 
\beq
A_0 > 0, \quad 3A_0A_2 - A_1^2 > 0. 
\label{pos}
\eeq
If the conditions (\ref{pos}) were not met, (\ref{lip}) would 
not describe a closed curve, but instead extend to infinity. This
is unphysical, since it would imply that the shock has spread 
an infinite distance. As the ellipse (\ref{lip}) is inserted 
into (\ref{gen_sol1}), one obtains a closed lip-shaped region, an example
of which is shown in Fig.~\ref{fig:lip}.  In the case $A_1=A_3=0$, namely for initial data with a symmetry with respect to reflection on the $y$-axis
one has
\[
\xi=\pm\dfrac{2}{3\sqrt{3A_0}}\left(\dfrac{\gamma+1}{2}-A_2\eta^2\right)^{\frac{3}{2}}.
\]
  The lip describes how the 
overturned region (and thus the shock) spreads in space, and corresponds
to similar results found in (\cite{EHHS14}) and (\cite{GKE16}). 

To find the position of the shock, we transform the solution 
(\ref{gen_sol1}) to a form equivalent to that of the one-dimensional
case. Namely, we can introduce shifted variables 
\beq
\bar{\xi} = \xi - \xi_s(\eta), \quad 
\bar{U} = U - U_s(\eta), 
\label{shift}
\eeq
so that (\ref{gen_sol1}) becomes 
\beq
\bar{\xi} = -A_0 \bar{U}\left(\bar{U}^2 - \Delta^2(\eta)\right). 
\label{gen_sol_trans}
\eeq
Comparing coefficients, we obtain 
\beqa
\label{coeff1}
&& U_s = -\frac{A_1}{3 A_0}\eta, \quad
\xi_s = -\frac{A_1(\gamma+1)}{6 A_0}\eta + 
\frac{9A_0A_1A_2 - 2A_1^3-27A_0^2A_3}{27 A_0^2}\eta^3, \\
\label{coeff2}
&& \Delta = \sqrt{\frac{\gamma+1}{2 A_0} + 
\frac{A_1^2-3A_0A_2}{3 A_0^2}\eta^2}. 
\eeqa
The lateral width of the shock is determined from the condition that 
$\Delta = 0$, and thus 
\beq
\eta_{\pm} = \pm\sqrt{\frac{3A_0(\gamma+1)}{6A_0A_2-2A_1^2}},
\label{width}
\eeq
where (\ref{pos}) guarantees that this is well-defined. 
Clearly, in real space the width of the shock increases like 
$|t'|^{1/2}$.

Having written the profile in the form of a simple s-curve
(\ref{gen_sol_trans}), if follows from symmetry that the shock 
must be at $\bar{\xi}=0$, so that the shock position is at 
$\xi_s(\eta)$. This is the dashed line plotted in Fig.~\ref{fig:lip}. 
The line $\bar{\xi}=0$ intersects (\ref{gen_sol_trans}) at 
$\bar{U}=\pm\Delta$, and so the velocities at the front and back 
of the shock are $U_1=U_s-\Delta$ and $U_2=U_s+\Delta$, respectively, 
so that the size of the jump is $2\Delta$. 

In real space the shock position is at 
\beq
\begin{split}
x_s'& = -c_0t' + By'^2 + |t'|^{3/2}\xi_s(\eta) \\
&= By'^2
-\left(c_0 - \frac{A_1(\gamma+1)}{6 A_0}y'\right) t' +
\frac{9A_0A_1A_2 - 2A_1^3-27A_0^2A_3}{27 A_0^2}y'^3,
\label{xs}
\end{split}
\eeq
so that the shock speed in the $x$-direction is
\beq
u_s = c_0 - \frac{A_1(\gamma+1)}{6 A_0}y'; 
\label{us}
\eeq
the speed in the $y$-direction is of lower order. 

To confirm that (\ref{us}) is in agreement with the Rankine-Hugoniot
conditions at the shock, note that according to (\ref{gen_sol_trans}),
the fluid velocities at the back and the front of the shock are
\beq
u_{1/2} = |t'|^{1/2}\left(U_s \pm \Delta\right).
\label{ulr}
\eeq
Using mass conservation, the shock velocity is (\cite{LL84a})
\[
u_s = \frac{\rho_1u_1-\rho_2u_2}{\rho_1-\rho_2},
\]
which to leading order can be written as 
\[
u_s = \left.\frac{\partial \rho u}{\partial \rho}\right|_{u = (u_1+u_2)/2}
= \frac{u_1+u_2}{2} + 
\rho\left.\left(\frac{\partial u}{\partial \rho}\right)^{-1}
\right|_{u = (u_1+u_2)/2} = c_0 + \frac{\gamma+1}{4}\left(u_1+u_2\right),
\]
where in the last step we used (\ref{simple}). Combining this 
with (\ref{ulr}) and the expression for $U_s$, one indeed recovers 
(\ref{us}). It is straightforward to check that the other 
Rankine-Hugoniot conditions are satisfied identically to leading order. 

\section{Numerical simulation}
\label{sec:num}
We test the results of the preceding sections by direct numerical 
simulation of the Euler equation.
Starting from a smooth initial condition for the density and 
the velocity, a shock develops. Our aim is to compare to the 
similarity profile (\ref{gen_sol1}), both before and after shock
formation, and to confirm the self-similar properties of the solution,
as described by (\ref{similarity_phi}),(\ref{similarity_density}). 
We have seen in earlier work (\cite{GKE16}) that it is much easier to 
test self-similar properties of profiles {\it after} the singularity,
where they have more structure. We will pursue this idea but with the 
additional twist that we use (\ref{coefficients}) {\it before} the 
singularity to calculate the coefficients $A_0-A_3$, which determines 
the self-similar solution completely. We are then able to predict 
profiles after the singularity without any adjustable parameters. 

We begin with the initial condition 
\beq
\rho(x,y,0) = 0.2 + e^{-4x^4-4y^2}, \quad
{\bf v}(x,y,0) = 0,
\label{init}
\eeq
which corresponds to a localized high-density, high-pressure region 
(as if generated by an explosion), starting from rest. We choose 
the adiabatic exponent of air $\gamma = 7/5$, and $A = \gamma$ in 
the ideal gas law (\ref{adiabatic}). The initial condition was chosen such 
that gradients are steeper in the $x$-direction, so that a shock first
occurs on the $x$-axis. Further, the solution is symmetric about the 
$x$-axis, so that the coefficients $A_1$ and $A_3$ in the self-similar
solution (\ref{gen_sol1}) vanish. This makes it much easier to spot 
the singularity; in particular, the $x$ and $y$ axes of the simulation 
are the same as those defined by the gradients of the density, in 
that $\partial \rho/\partial y=0$ is satisfied. 

For times $t<t_0$ we use both a finite difference scheme and a Fourier
pseudo-spectral method to solve the equations. In the finite-difference
scheme, the equations are written as in (\ref{adiabatic})-(\ref{Euler}), 
and are discretized in space using fourth order finite differences on a uniform 
mesh in the numerical domain $[0,2]\times[0,2]$. Mirror symmetry is
applied at $x=0$ and at $y=0$, while outflow conditions (vanishing 
derivatives of all variables) are applied at $x=2$ and at $y=2$. 
An explicit second order Runge-Kutta scheme is used to advance the 
solution in time. We used $2000\times 2000$ points in space and a 
fixed time step of $\Delta t = 1.25\cdot 10^{-4}$. 

For the pseudospectral method (\cite{CHQZ1}), \eqref{continuity} and
\eqref{Ber} are set in a $[-2\pi,2\pi]\times[-2\pi,2\pi]$ box with 
periodic boundary conditions, with an equispaced collocation grid of 
resolution $2^{14}\times 2^{14}$. The time discretization is obtained 
by means of a standard fourth order explicit Runge-Kutta scheme with 
$\delta t = 5\cdot 10^{-5}$. To remove aliasing errors, we adopt a filtering as 
described in~\cite{hou2009}, whereby Fourier coefficients are multiplied 
by the exponential function
\begin{equation}
  \sigma(k)=\exp(-36(|k|/N)^{36}),
  \label{eeq:def_fourier_filter}
\end{equation}
where $N=2^{14}$ is the number of Fourier modes in each direction.

As the singularity is approached, steepening gradients require 
higher and higher Fourier modes to represent profiles accurately. 
To guarantee sufficient resolution as the profiles steepen,
we inspect the magnitude of the Fourier coefficients at each time step. 
As long as all Fourier modes with magnitude higher than the machine epsilon 
($10^{-12}$) are represented, the approximation is deemed acceptable; 
if this is no longer the case for a given resolution, we stop the 
simulation. 

\begin{figure}[htbp]
\centering
\includegraphics[width=0.45\textwidth]{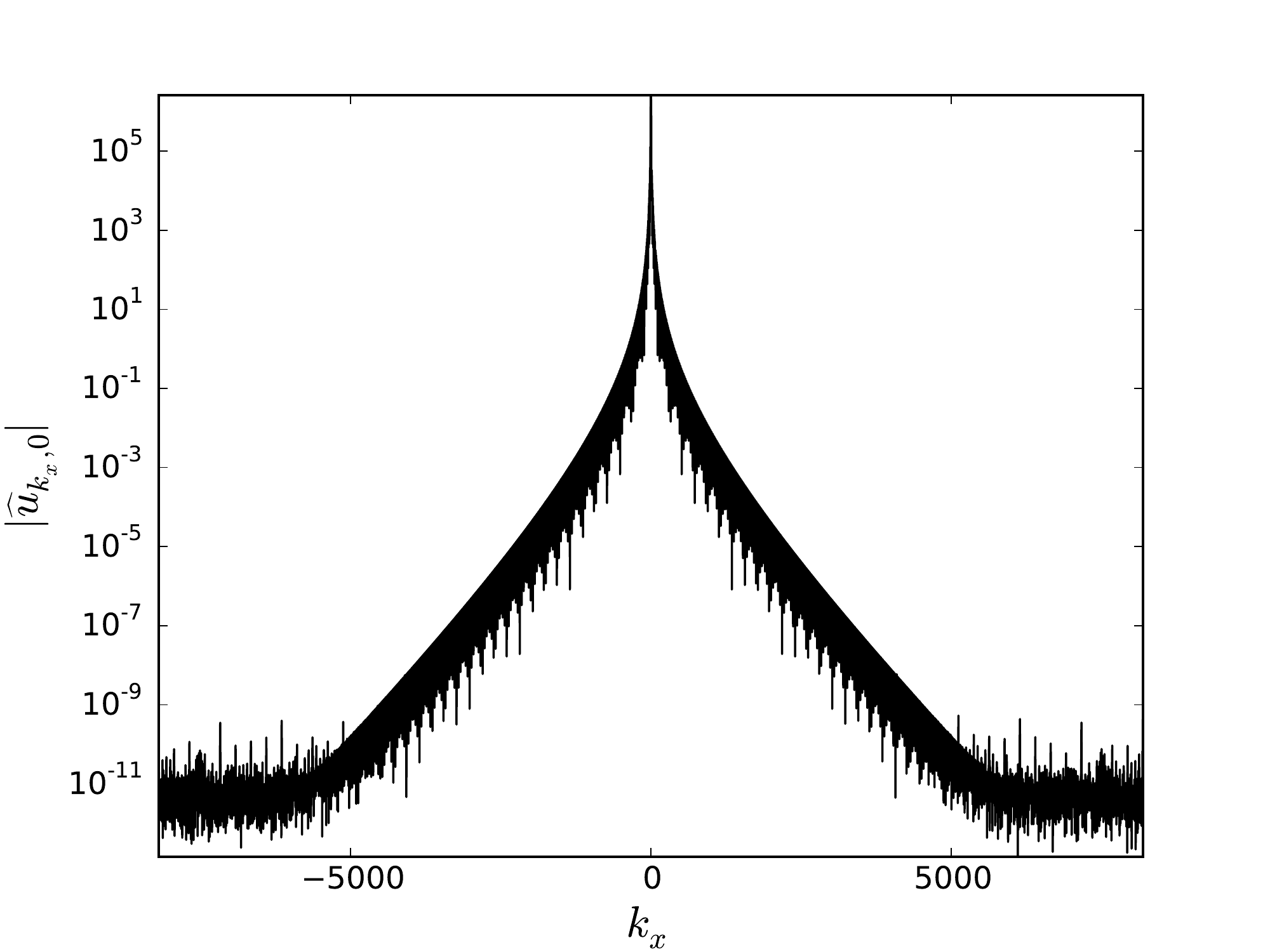}
\includegraphics[width=0.45\textwidth]{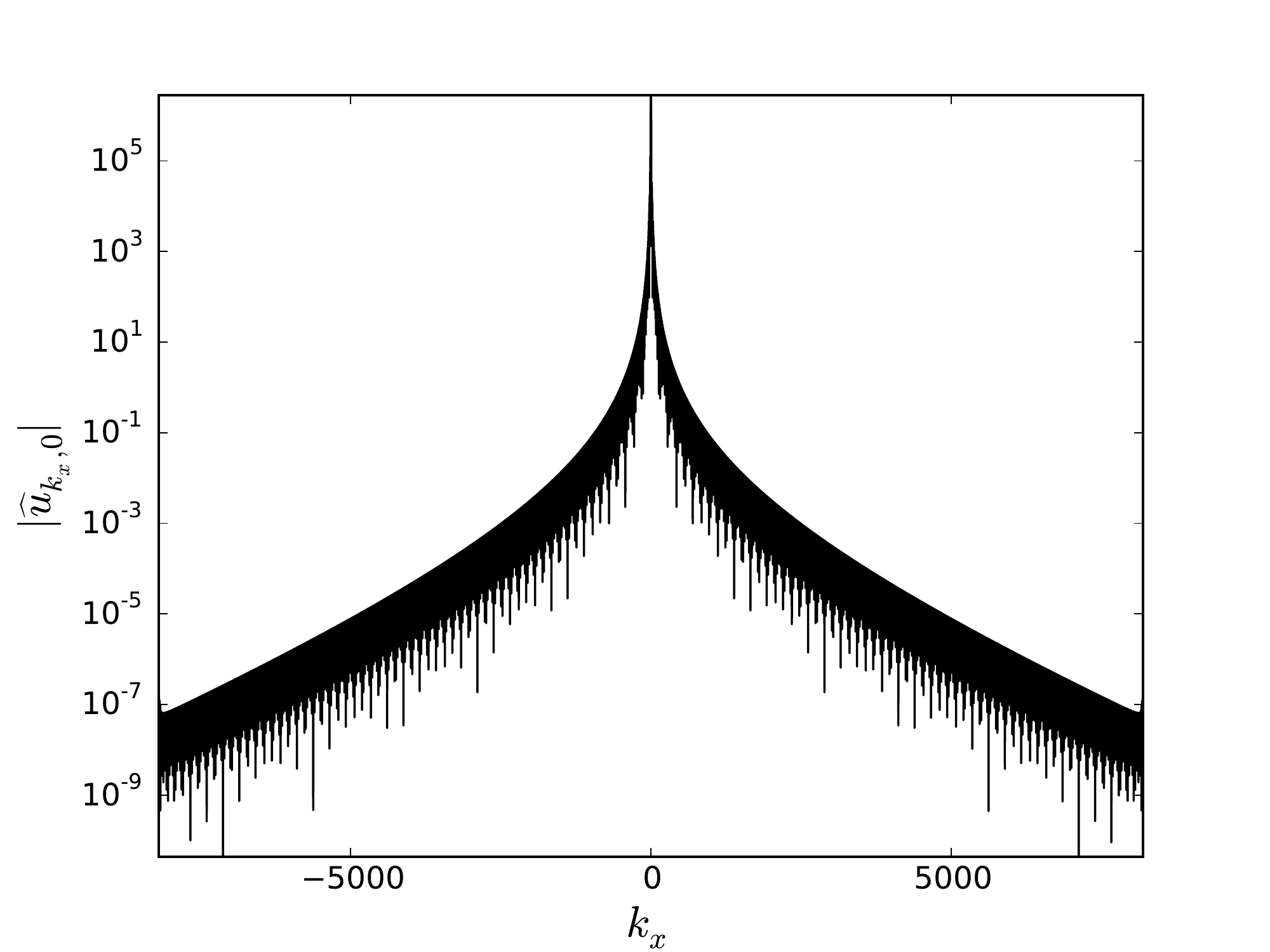}
\caption{Magnitude of Fourier coefficients of the numerical 
solution as function of the $x$-wavenumber at the origin of the 
$y$-wavenumber axis. At $t=0.45$ (left) there is sufficient resolution,
while at $t=0.48$ (right) the spectrum has been rejected. 
}
\label{fig:phihat}
\end{figure}

As an example, in Fig.~\ref{fig:phihat} we report the spectrum for an 
acceptable solution (at time $t=0.45$) on the left, and for a rejected 
solution (at time $t=0.48$) on the right. On the left, Fourier amplitudes 
plateau to the smallest representable number $10^{-12}$, and thus the 
simulation can be trusted to within the arithmetic precision of the 
calculation, while on the right this is no longer the case. 
On the basis of this, we continue the pseudospectral calculation up 
to $t=0.46$, and perform a least squares interpolation of this part of 
the solution to extrapolate to the critical time. 
 
When a shock appears,  we need to use  a finite difference method that 
remains stable even in the presence of jumps of the 
hydrodynamic fields. To this end, the equations are written
in conservative form, where the fluxes are computed using the 
second-order-in-space central-upwind scheme (see e.g. Section 3.1 of 
(\cite{KL02}) with slope limiting (\cite{vL79}).
In addition to $\rho$ and the mass flux ${\bf j} = \rho{\bf v}$,
the method uses the internal energy 
$e = \rho{\bf v}^2/2 + p/(\gamma-1)$ as an additional 
variable. The energy follows the conservation equation 
\beq
\frac{\partial e}{\partial t} + \nabla\cdot({\bf v}(e+p)) = 0,
\label{energy}
\eeq
while (\ref{continuity}) and (\ref{Euler}) are solved as before,
but in conservative form. Energy dissipation occurs within a tiny region 
around the shock, where entropy is created. This method remains stable even 
if the shock is not resolved, effectively modeling non-classical solutions 
to the Euler equation, which satisfy the Rankine-Hugoniot conditions. 
Time integration is performed with a generic variable time-step 
predictor-corrector scheme. 

This numerical scheme is implemented using the ``Basilisk'' software, developed 
by S. Popinet. It uses Quadtrees (\cite{Pop11}) to allow efficient adaptive 
grid refinement in the region where the gradient of the density or of 
the velocity becomes large. Linear refinement is used on the trees, 
so that reconstructed values also use slope limiting. 
The numerical domain is $[-2,2]\times[-2,2]$, i.e. symmetry conditions 
are  not applied in this case, and it is discretized using 
$2^{10}\times 2^{10}$ points initially. The resolution is adapted at each 
time step according to the (wavelet-estimated) discretization error of 
the density field. Whenever the discretization error is larger than 
$5\times10^{-3}$, the mesh is refined, down to a prescribed maximum quadtree 
level. Several simulations have been carried out by varying the maximum 
level of refinement from 10 to 18.  

\begin{figure}
\centering
   \includegraphics[width=0.8\textwidth]{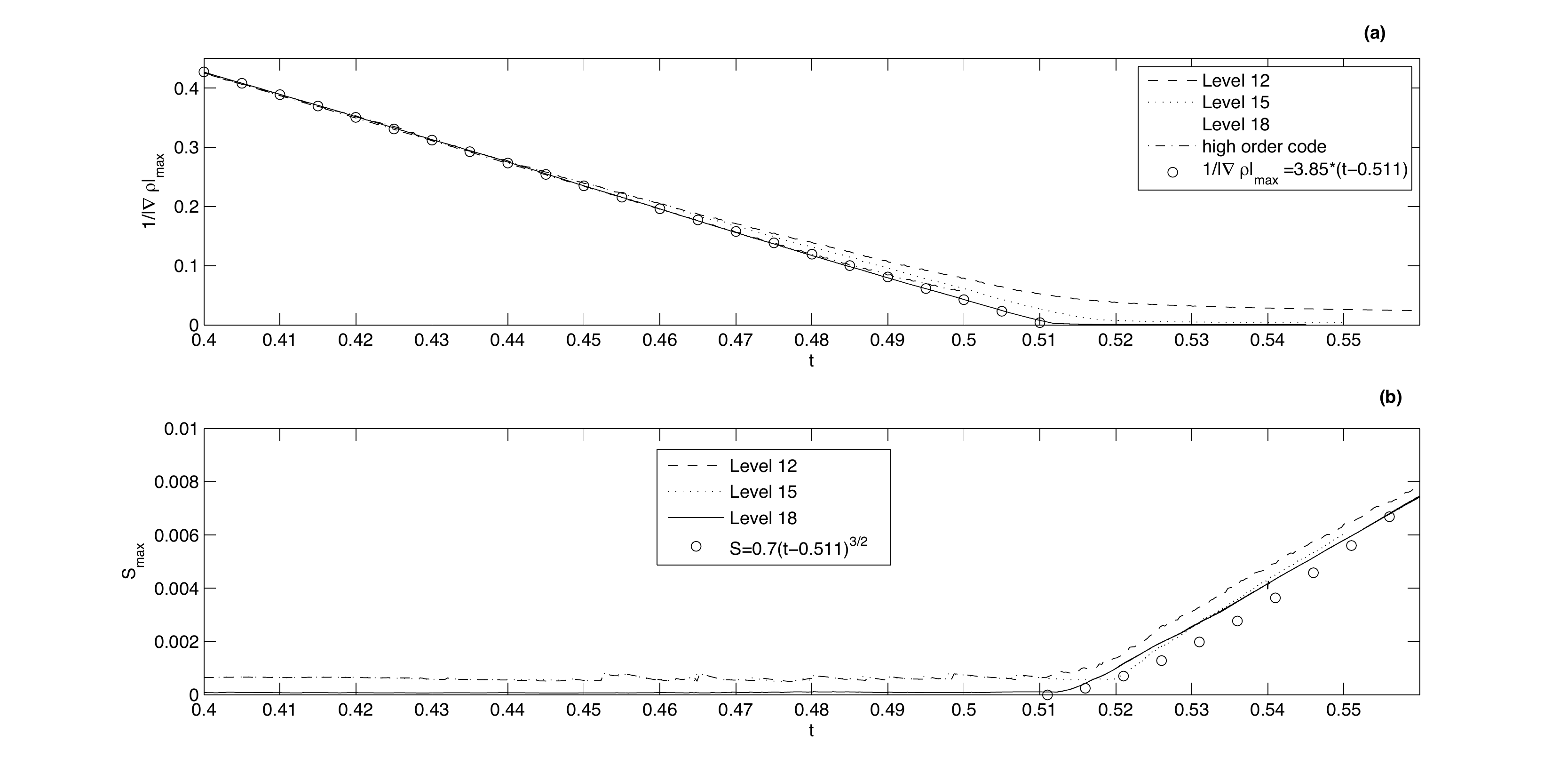} 
    \includegraphics[width=0.8\textwidth]{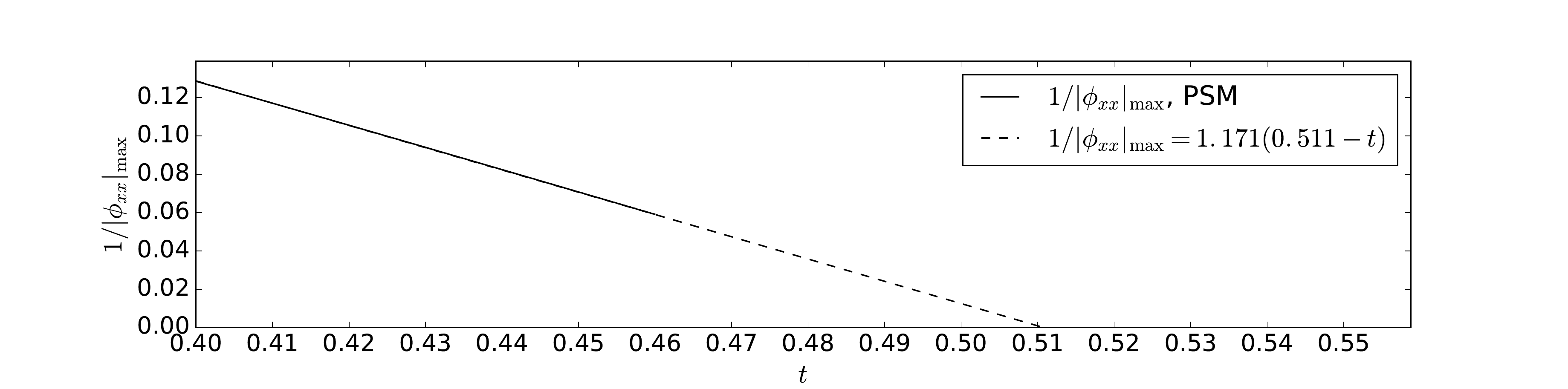}
  \caption{The inverse of the maximum of the density gradient (top), 
and the maximum of the entropy (center) over the whole domain 
as function of time, for three different levels of resolution, 
using the Basilisk code for the compressible Euler equation. 
The dot-dashed line is the result of the fourth-order code,
up to $t=0.5$. By fitting to the predicted linear law (\ref{max_grad}), we get 
an accurate prediction for the singularity time $t_0=0.511$. 
In the bottom panel, the inverse of the maximum of $\phi_{xx}$, 
obtained using the pseudospectral method, is plotted as a function of 
time. A linear fit leads $t_0=0.512$. 
   }
\label{fig:blowup}
\end{figure}

To locate the singularity, we look at the maximum gradient of 
the density $\rho$ and the velocity field $u$, which is in the $x$-direction:
\[
\frac{\partial \rho}{\partial x} = \frac{\rho_0}{c_0}|t'|^{-1}U_\xi,
\quad u_x=|t'|^{-1}U_\xi.
\]
According to the similarity solution (\ref{gen_sol1}), the minimum
$\xi_U=-(\gamma+1)/2$ is at $U=\eta = 0$, and hence 
\beq
\left|\grad\rho\right|_{max}^{-1} = \frac{c_0}{\rho_0}\frac{1+\gamma}{2}|t'|,
\quad \left|u_x\right|_{max}^{-1} =\frac{1+\gamma}{2}|t'|.
\label{max_grad}
\eeq
The predicted linear dependence for $t<t_0$ is confirmed in 
Fig.~\ref{fig:blowup} (top and bottom).  The quantity 
$\left|\grad\rho\right|_{max}^{-1} $ is computed using the conservative
scheme, since it allows us to go up to the singularity and beyond.
Close to the singularity, the maximum gradient of the density
crosses over to a finite value, as the scheme can no longer resolve the 
steepest gradient. As the resolution is increased, the linear behavior 
continues to smaller values. From a linear fit of the inverse of $\rho_x$ 
to the highest resolution data (circles), we find $t_0 = 0.511$, which is 
our most accurate estimate of the singularity time, since it is based 
on a simulation which continues up to shock formation and beyond. 
Using (\ref{max_grad}), the prefactor of the linear fit is $4.01$, 
in reasonable agreement with the fitted value of $3.85$. The linear 
fit also agrees very well with the result of the fourth order 
finite-difference code before the singularity. 

To confirm that the velocity component $u$ blows up in the same way as 
$\rho$, we use the pseudospectral method to also calculate 
$\left|u_x\right|_{max}^{-1} $. A linear fit to
$\left|u_x\right|_{max}^{-1} =|\phi_{xx}|_{max}^{-1}$ gives a singularity 
time of $0.512$, in good agreement with the result of the finite difference 
scheme. The prefactor of the linear fit is $1.171$, again in good
agreement with the  theoretical value $(\gamma+1)/2=1.20$ 
(see bottom of Fig.~\ref{fig:blowup}).

Using the location of the maximum gradient of $\rho$ at $t_0$, we obtain 
${\bf x}_0= (1.4052,0)$ as the position of the singularity. At this point,
the velocity ${\bf v}_0 = (0.2769,0)$, the density $\rho_0=0.2731$, 
and $c_0=  0.9127$. From now on, we will report all results in a frame 
of reference which moves with ${\bf v}_0$, and relative to ${\bf x}_0$.
The middle graph of Fig.~\ref{fig:blowup} shows the maximum entropy, 
which starts to grow exactly at the time of shock formation $t_0$. 
The growth is consistent with a fit based on (\ref{s_jump}), which 
would predict the maximum entropy to grow like $t'^{3/2}$. However, 
our results are not sufficiently accurate to distinguish this from a 
linear behavior. 

\begin{figure}
\centering
\includegraphics[width=0.49\textwidth]{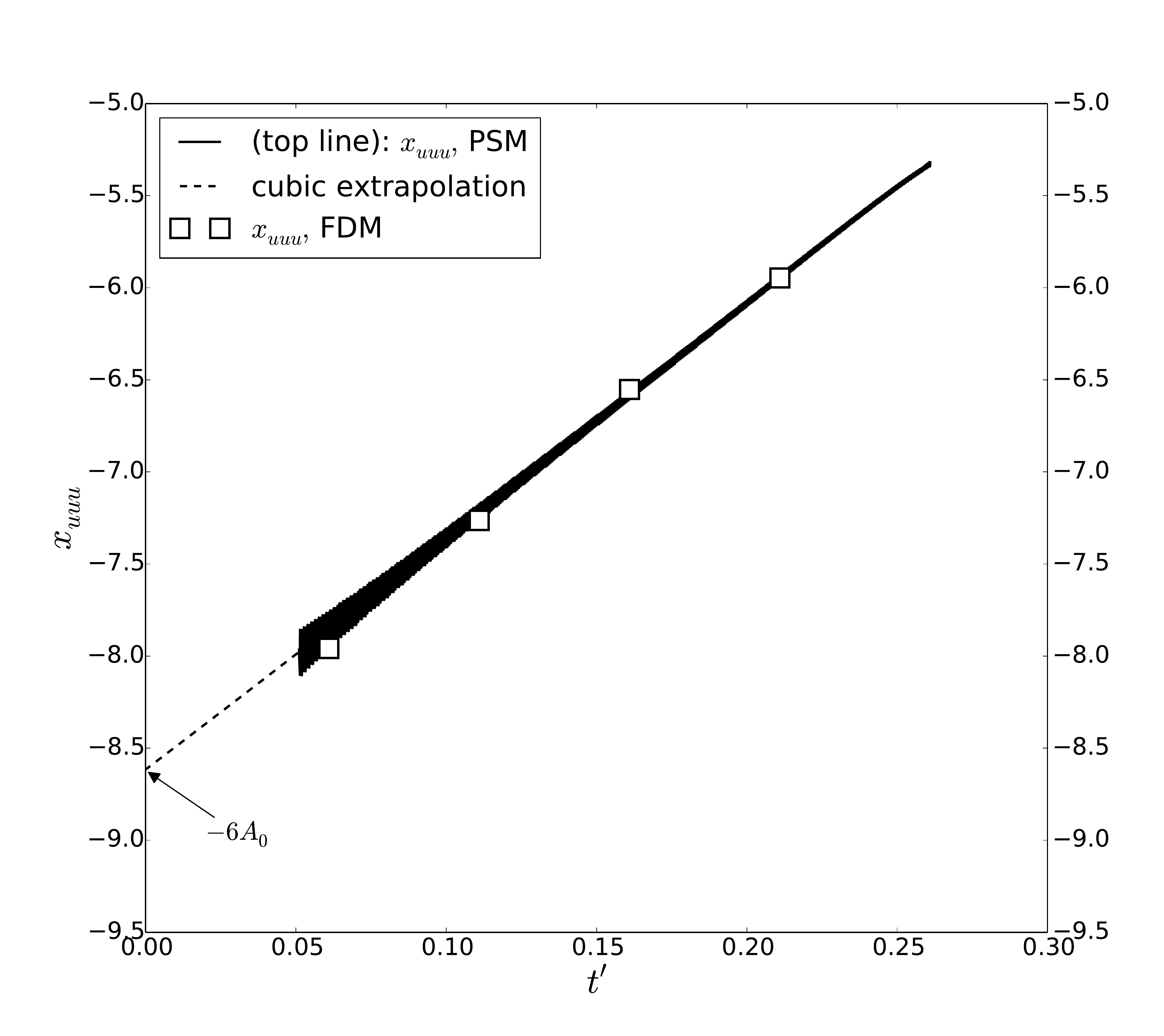} 
\includegraphics[width=0.49\textwidth]{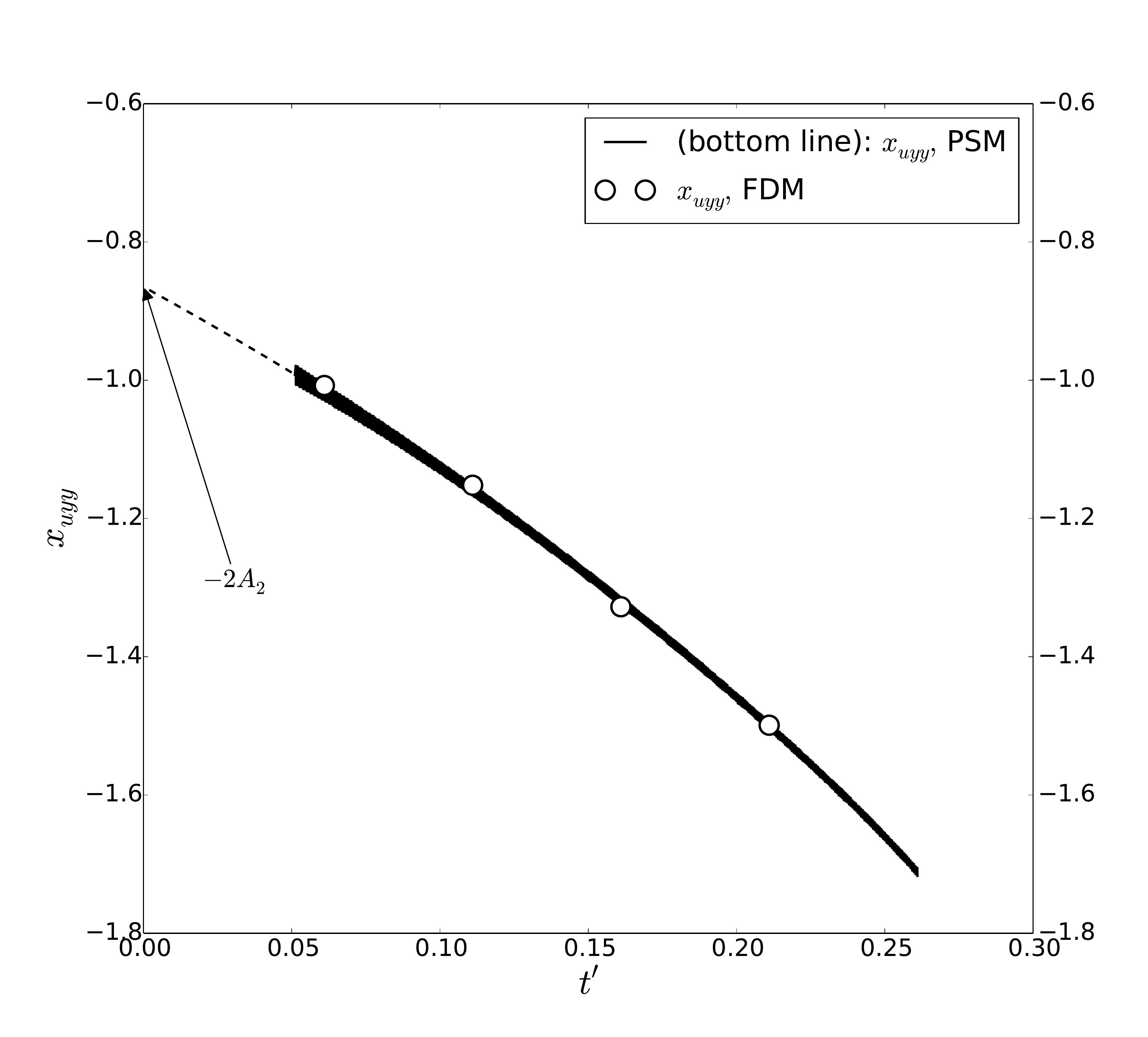} 
\caption{The derivatives $x_{uuu}$ (left) 
and $x_{uyy}$ (right)  as function of $t'$ for the initial conditions 
(\ref{init}), evaluated at the maximum of the pressure, using the 
PseudoSpectral Method (PSM) and the Finite Difference Method (FDM). 
The values of$x_{uuu}$  and $x_{uyy}$ are plotted  before they 
start to oscillate strongly for $t'<0.05$. The coefficients 
$A_0$ and $A_2$ are found by extrapolating to $t'=0$.
}
\label{fig:Ai}
\end{figure}

\begin{figure}
\floatbox[{\capbeside\thisfloatsetup{capbesideposition={right,top},
capbesidewidth=4cm}}]{figure}[\FBwidth]
{ \caption{Plot of the  values of $u_{yy}t'$ for several values 
of $t'$ using PSM and FDM. The  coefficient $B$
is found from (\ref{B}) by extrapolating to $t'=0$. 
 }
\label{figureB}}
{\includegraphics[width=5cm]{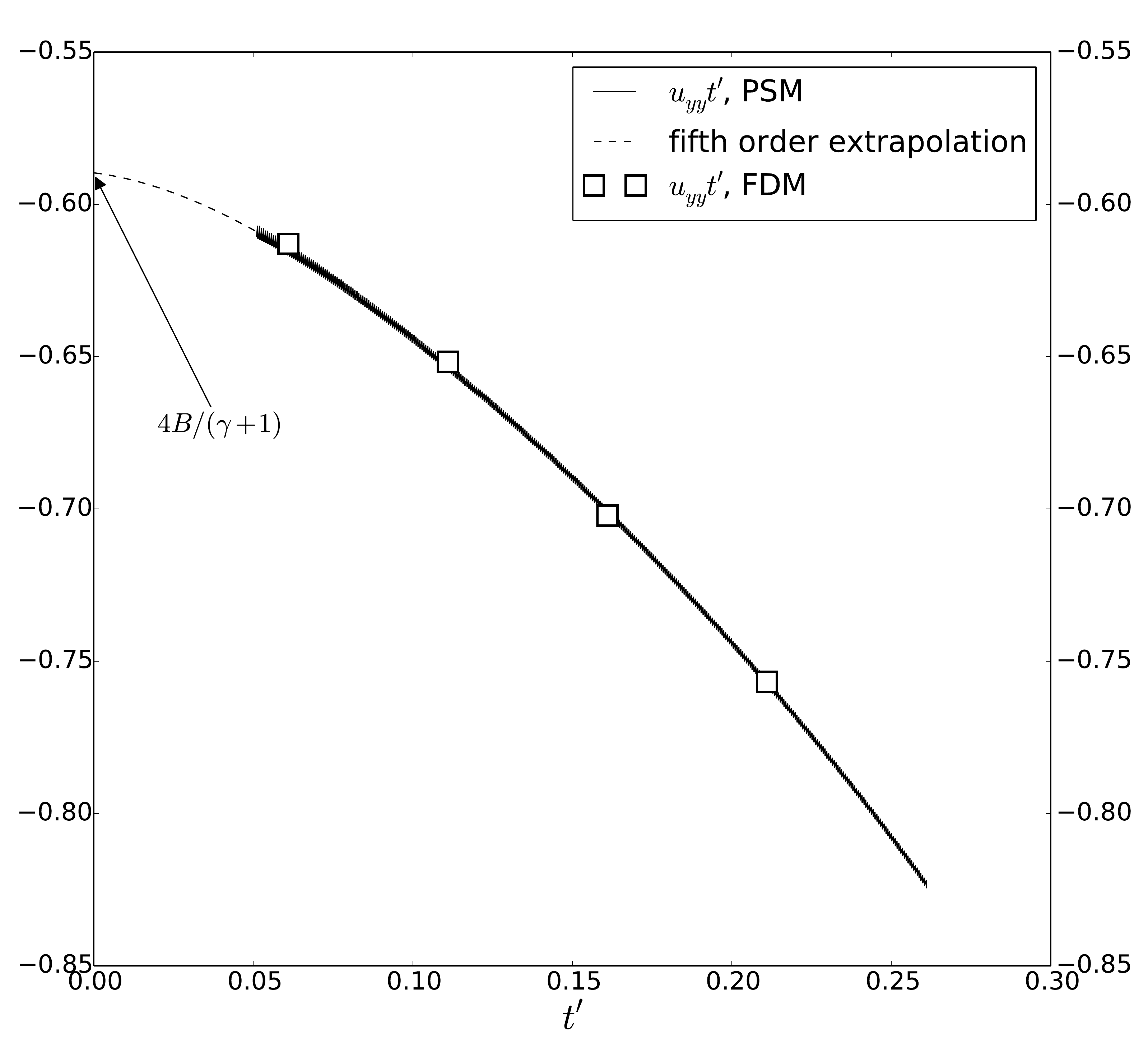}}
\end{figure}

\begin{table}
\centering
\begin{center}
\begin{tabular}{|c|c|c|c|}
\hline
   & FDM linear extrapolation & PSM&extrapolation \\
     \hline
  $A_0$ & $1.5508$               & $1.43614$     &cubic                  \\
  \hline
  $A_2$ & $0.3328$                & $0.43242$       &cubic                \\
  \hline
  $B$   &  -0.33          & $-0.35377$         &quintic             \\
\hline\end{tabular}
\end{center}
\medskip
\caption{The parameters of the similarity solution as determined 
numerically from $t<t_0$, extrapolating to $t_0 = 0.511$, 
using the finite difference and pseudospectral methods.
\label{tab:cmp_results}
}
\end{table}

We are now in a position to calculate the constant $B$ 
(cf. (\ref{similarity_phi}) as well as the coefficients $A_i$ 
which appear in the similarity solution (\ref{gen_sol1}); 
by symmetry, $A_1=A_3=0$. For the latter, We use (\ref{coefficients}), 
aiming to evaluate the right-hand sides as close to the singularity as 
possible. As illustrated in Figs.~\ref{fig:Ai} and \ref{figureB}, 
and recorded in Table~\ref{tab:cmp_results}), we use the results of 
both the the finite difference method (FDM) and our pseudo-spectral 
method (PSM) to extrapolate to $t_0$. 
As seen from (\ref{xuuu})-(\ref{xyyy}), the numerical approximation 
for the third derivatives will loose resolution eventually, since for example 
$u_x$ blows up at the singularity, and cancellation errors become large. 
In calculating $x_{uuu}$ and $x_{uyy}$, we use that odd derivatives 
with respect to $y$ vanish on account of symmetry. We then evaluate 
$x_{uuu}$ and $x_{uyy}$ at the maximum of the pressure, and plot the result 
as a function of time see, Fig.~\ref{fig:Ai}. We use linear and cubic 
approximations to extrapolate $x_{uuu}$ and $x_{uyy}$  to $t'=0$, from 
which $A_0$ and $A_2$ are calculated 
using (\ref{coefficients}) (see Table~\ref{tab:cmp_results}).
The coefficient $B$ (cf. (\ref{similarity_phi})),
is found from (\ref{B}) by extrapolating to $t=t_0$, using both linear 
and quintic approximations (see Fig.~\ref{figureB} and 
Table~\ref{tab:cmp_results}). As seen in Table~\ref{tab:cmp_results}), 
the numerical values for the coefficients, obtained by different methods,
are in good agreement 
\begin{figure}
\centering
\includegraphics[width=0.8\textwidth]{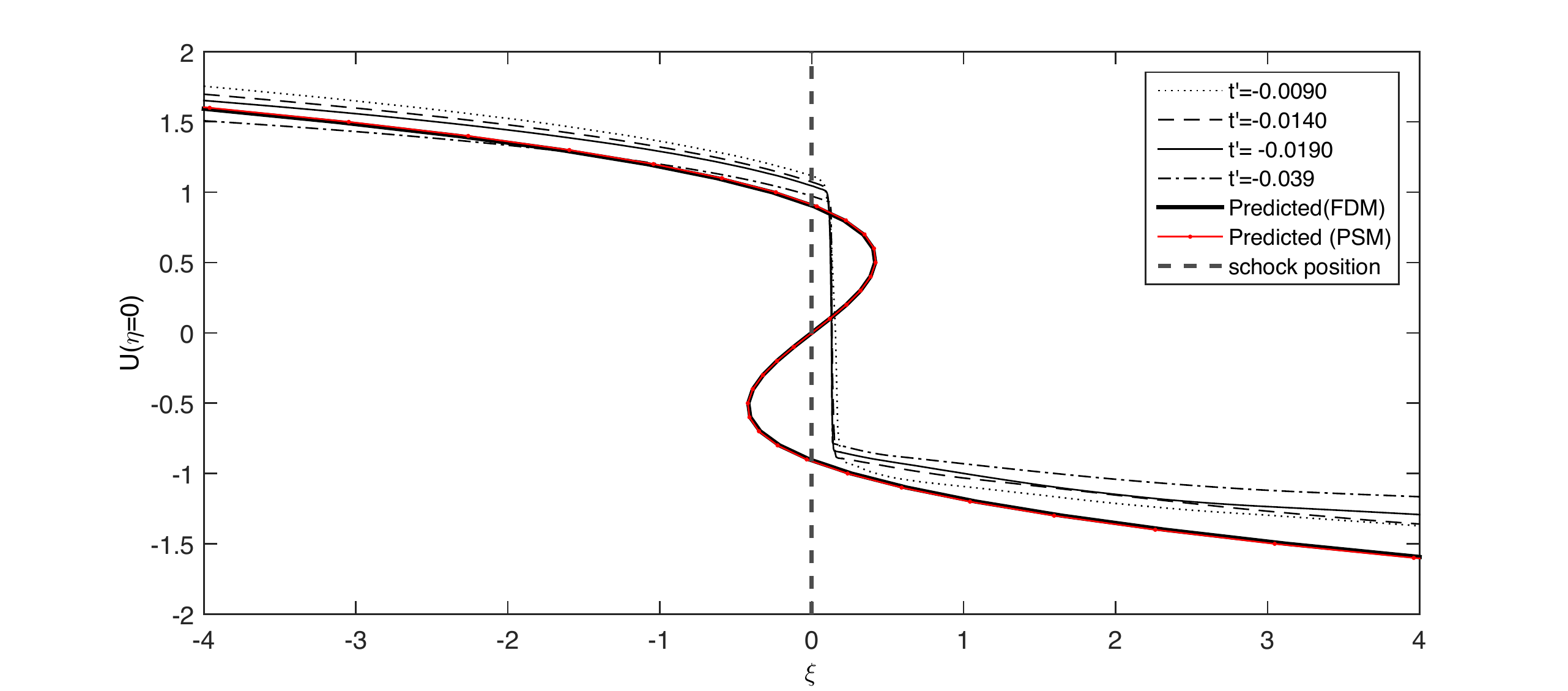} 
\caption{Velocity profiles $u(x,y=0,t)$, rescaled according to 
(\ref{U_scal}), for $t'$ as given. The heavy line as well as the red 
line with dots correspond to (\ref{gen_sol1}) for $\eta = 0$, using the 
two sets of coefficients. The jump is predicted to be at $\xi=0$ 
(heavy dashed line). 
}
 \label{fig:U_profile}
   \end{figure}

From (\ref{similarity_phi}), the velocity field $u$ in the direction 
of propagation along the axis of symmetry is 
\beq
u - u_0 = |t'|^{1/2}U\left(\frac{x' + c_0t'}{|t'|^{3/2}},0\right). 
\label{U_scal}
\eeq
In Fig.~\ref{fig:U_profile}, velocity profiles have been rescaled 
according to (\ref{U_scal}), and superimposed for the times shown. 
Note that no adjustable parameter was used to achieve the collapse, 
which requires accurate estimates for $x_0$ and $u_0$, as well as the 
speed of sound $c_0$. Theoretical predictions for the profile 
(\ref{gen_sol1}), based on the coefficients from 
Table~\ref{tab:cmp_results}, are shown as the heavy black line (FDM),
and the red line with dots (PSM), giving almost identical results. 
The theoretical prediction for the shock position is a jump inserted 
at $\xi = 0$. Although there is no adjustable parameter in the comparison, 
profiles collapse very well over a wide range of $t'$ values, and agree 
with the theoretical prediction, based on an independent determination of the 
free parameters.

\begin{figure}
\centering
   \includegraphics[width=0.8\textwidth]{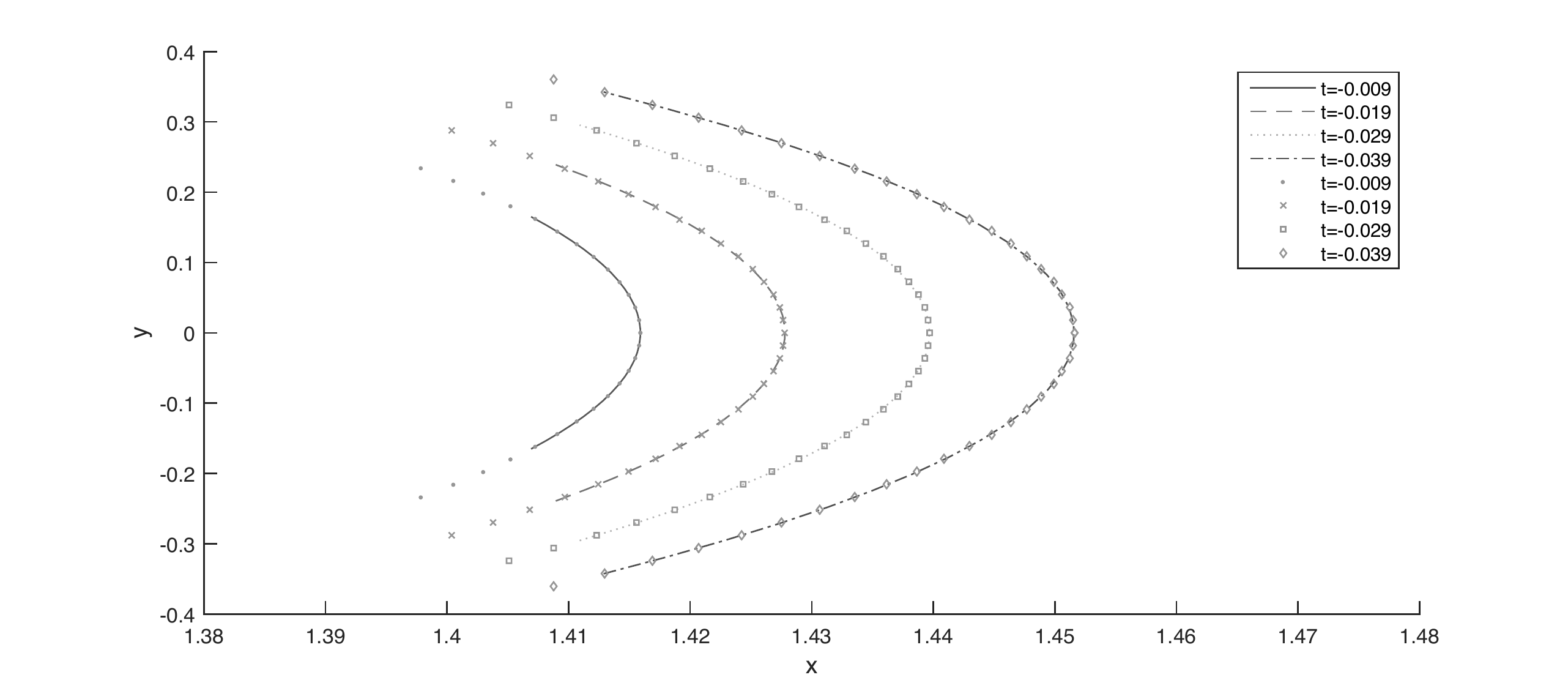} 
  \caption{
The position of the shock front as function of time, as determined 
from the maximum of the density gradient. This is superimposed with 
the theoretical prediction (\ref{xs}), with $B$ as given in
Table~\ref{tab:cmp_results}.}
\label{fig:shape}
\end{figure}
\begin{figure}
\centering
   \includegraphics[width=0.8\textwidth]{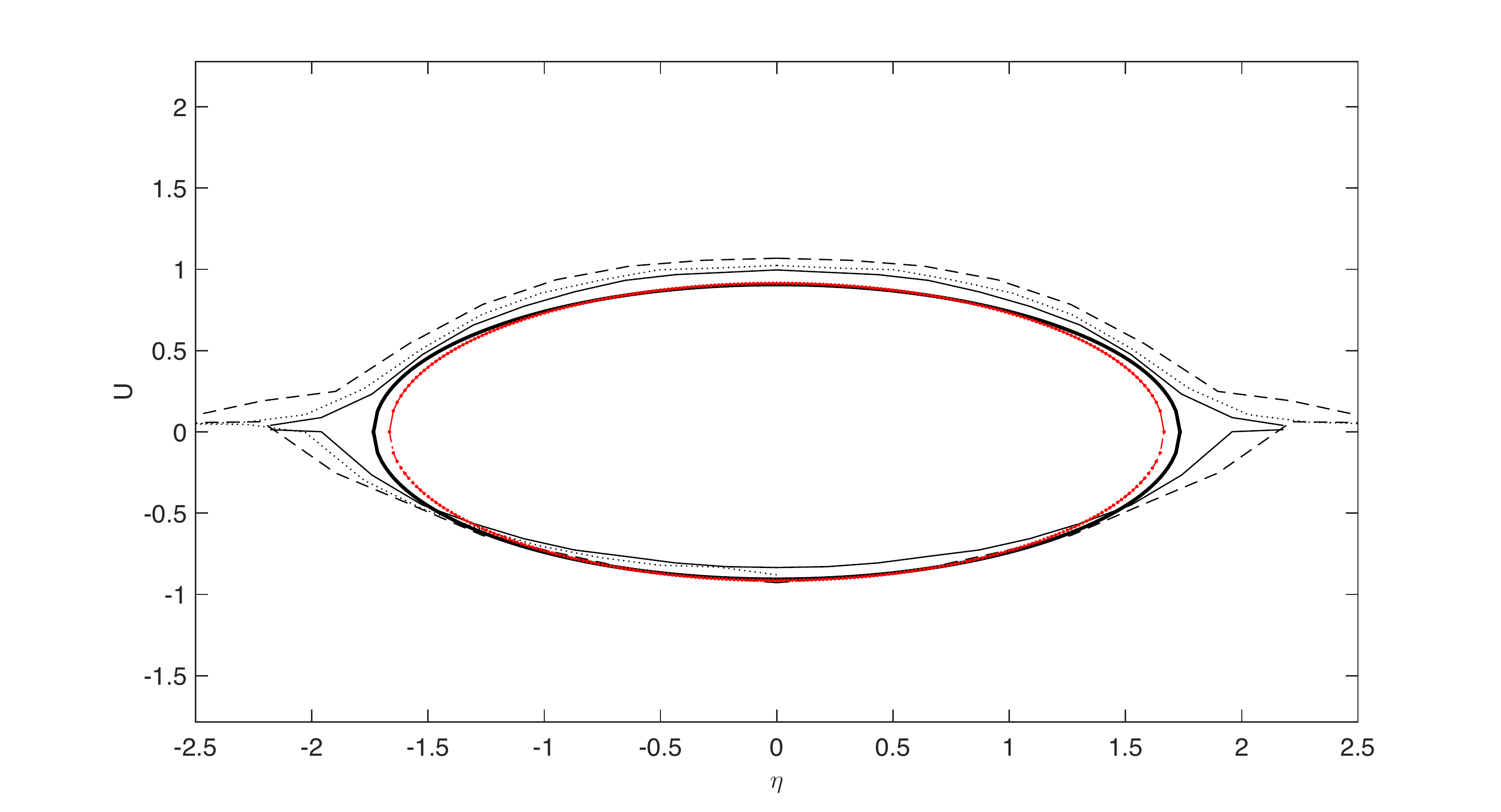} 
  \caption{
The rescaled values of the velocity field $U_{1,2}$ at 
the front and back of the shock, written as function of the similarity 
variable $\eta$. Numerical results are for $t' = -0.009$ (dashed line), 
$t' = -0.014$ (dotted line), and $t' = -0.039$ (solid line). The 
heavy solid line and the red line with dots are the theoretical predictions 
based on (\ref{coeff2}). 
}
\label{fig:width}
\end{figure}
In Figs.~\ref{fig:shape} and \ref{fig:width} 
we test the spatial structure of the shock and how it spreads in time. 
First we show the position of the shock in real space 
(cf. Fig.~\ref{fig:shape}), as determined from the maximum gradient of the 
density. This procedure does not determine where the profile has a true jump, 
so we also have to calculate the height of the jump, for which we use a 
procedure described below, as applied to the $x$-velocity $u$. We thus see 
the lateral spreading of the shock as it propagates forward. This is 
compared to the theoretical prediction (\ref{width}),(\ref{xs}), with 
which excellent agreement is found. This confirms that the width of the 
shock spreads like $|t'|^{1/2}$, with a prefactor (\ref{width}) determined 
from the coefficients $A_i$. It also shows that the shape of the shock front
is as predicted by (\ref{xs}). 

To look at the structure of the shock in more detail, we consider
the velocity in the front and back of the shock, 
$U_{1/2}=U_s\mp\Delta$, as given by (\ref{coeff1}),(\ref{coeff2}). 
This prediction is shown as the heavy black line and the red line 
with dots in Fig.~\ref{fig:width}, with coefficients 
determined before the singularity. The values of $u_1$ and $u_2$ 
are determined numerically from slices such as those shown in 
Fig.~\ref{fig:U_profile}, but for a range of $y$ values, until the shock 
disappears. Similarity functions are found by rescaling according to 
$U_{1/2} = \left(u_{1/2}-u_0\right)/|t'|^{1/2}$ and 
$\eta = y'/|t'|^{1/2}$. 

Near the center of the shock, $u_1$ and $u_2$ are relatively easy to determine, 
by looking for a corner in profile, where it suddenly becomes vertical;
but as the shock becomes weaker near the edge, numerical viscosity
leads to a rounding of the jump, and values of $u_1$,$u_2$ can no longer 
be read off as easily. Instead, we adopt the following procedure:
first, the derivative of the profile has a sharp peak at the position
of the shock, which we take as the location of its minimum. Second, 
we fit a third order polynomial to both the upper and lower branches 
of the profile, away from the region directly at the shock where 
numerical viscosity is significant. Then $u_1$ is found from the 
intersection of the upper branch with a vertical line at the 
position of the shock, and $u_2$ from the lower branch. The result
of this procedure is shown for three different values of $t'$. 
Again, excellent collapse is found, as well as agreement with the
theoretical prediction, based on the two sets of coefficients.  

\section{Discussion}
In this manuscript we have derived the leading order behavior of the 
solution of the compressible two-dimensional isentropic Euler equation 
near the formation of its first singularity. We have obtained a self-similar 
structure for the local solution near the singularity, showing 
it captures the main features of the local behavior of the shock solution 
after singularity formation. In particular, we find scaling like 
$t^{1/2}$ along the orthogonal direction of propagation and scaling like 
$t^{3/2}$ along the direction of propagation. Furthermore, for a specific 
choice of initial data, we have compared the spatial structure of the shock 
with our theoretical predictions, finding good agreement.

It is a worthwhile exercise to extend our calculations to
three space dimensions, in which case there are two variables 
$y$ and $z$ in the direction transversal to the direction of propagation
$x$. Repeating essentially the same steps as before, this leads to a 
similarity profile similar to (\ref{gen_sol1}), but which contains all 
third-order terms in the variables $U$ and the two similarity variables for the 
transversal directions. 

Our similarity solution is in the form of an infinite series 
(\ref{similarity_phi}),(\ref{similarity_density}), of which we 
calculated the leading order contributions $\Phi(\xi,\eta)$ and 
$R(\xi,\eta)$. It would be interesting to pursue the calculation to the 
next order and beyond, in order to calculate the contributions of higher 
order like $\Phi_1(\xi,\eta)$ and $Q(\xi,\eta)$. This will affect the 
transversal velocity component $v$, while our focus has been on the component 
$u$ in the direction of propagation. 

\section*{Acknowledgments}
JE's work was supported by a Leverhulme Trust Research Project Grant.


\end{document}